\documentstyle[epsfig,mnras_cite,times]{mn}

\title{Weak gravitational lensing: reducing the contamination by intrinsic alignments}
\author[Catherine Heymans \& Alan Heavens]{Catherine Heymans$^{1}$\thanks{ceh@astro.ox.ac.uk}\&
Alan Heavens $^{2}$\thanks{afh@roe.ac.uk} \\
$^1$ University of Oxford, Astrophysics, Keble Road, Oxford, OX1 3RH, UK \\
$^2$ Institute for Astronomy, University of Edinburgh, Royal Observatory,
Blackford Hill, Edinburgh, EH9 3HJ, UK}

\newcommand{\be}{\begin{equation}}  \newcommand{\ee}{\end{equation}}
  \newcommand{\ba}{\begin{eqnarray}}
\newcommand{\ea}{\end{eqnarray}}  
\newcommand{\nn}{\nonumber\\}

\newcommand{\bk}{{\bf k}}

\newcommand{\bx}{{\bf x}}  

\newcommand{\k}{\bar \kappa_{\rm eff}}
\newcommand{\th}{\overrightarrow\theta}
\newcommand{\f}{f_K}

\newcommand{\bm}[1]{\mbox{\boldmath{$#1$}}}
\def\gs{\mathrel{\raise1.16pt\hbox{$>$}\kern-7.0pt %
\lower3.06pt\hbox{{$\scriptstyle \sim$}}}}         %
\def\ls{\mathrel{\raise1.16pt\hbox{$<$}\kern-7.0pt %
\lower3.06pt\hbox{{$\scriptstyle \sim$}}}}         %

\begin{document}

\maketitle

\begin{abstract}
Intrinsic alignments of galaxies can mimic to an extent the
effects of shear caused by weak gravitational lensing.  Previous
studies have shown that for shallow surveys with median redshifts
$z_m \sim 0.1$, the intrinsic alignment dominates the lensing
signal.  For deep surveys with $z_m \sim 1$, intrinsic alignments
are believed to be a significant contaminant of the lensing
signal, preventing high-precision measurements of the matter
power spectrum.  In this paper we show how distance information,
either spectroscopic or photometric redshifts, can be used to
down-weight nearby pairs in an optimised way, to reduce the
errors in the shear signal arising from intrinsic alignments.
Provided a conservatively large intrinsic alignment is assumed,
the optimised weights will essentially remove all traces of
contamination. For the Sloan spectroscopic galaxy sample,
residual shot noise continues to render it unsuitable for weak
lensing studies.  However, a dramatic improvement for the
slightly deeper Sloan photometric survey is found, whereby the
intrinsic contribution, at angular scales greater than 1
arcminute, is reduced from about 80 times the lensing signal to a
10\% effect.  For deeper surveys such as the COMBO-17 survey with
$z_m \sim 0.6$, the optimisation reduces the error from a largely
{\em systematic} $\sim 220$\% error at small angular scales to a
much smaller and largely {\em statistical} error of only 17\% of
the expected lensing signal. We therefore propose that future
weak lensing surveys be accompanied by the acquisition of
photometric redshifts, in order to remove fully the unknown
intrinsic alignment errors from weak lensing detections.
\end{abstract}

\begin{keywords}
cosmology: observations - gravitational lensing - large scale
structure, galaxies: formation
\end{keywords}

\title{Weak gravitational lensing: reducing the contamination by intrinsic alignments}
\section{Introduction}

Weak gravitational lensing by intervening large-scale structure
introduces a coherent distortion in faint galaxy images.  Several
independent surveys have measured this `cosmic shear' effect and are
now able estimate the bias parameter, b, \cite{HYG01,HvWGMY} and set
joint constraints on the values of the matter density parameter
$\Omega_{\rm m}$ and the amplitude of the matter power spectrum,
$\sigma_8$ \cite{BMRE,MLB02,HYG02,Maoli,Rhodes,vWb01}.

A potential limitation of this technique is the intrinsic correlation
of the ellipticities of nearby galaxies which can result from
gravitational interactions during galaxy formation.  This `tidal
torquing' leads to a net alignment of nearby galaxies which could
produce a spurious signal similar to that induced by weak
gravitational lensing.  A number of studies have investigated the
amplitude of this intrinsic alignment effect with some broad
agreement, (\pcite{HRH00}, hereafter HRH,
\pcite{BTHD02,CKB01,CNPT01,CM00,HZ02,LP01,Jing,Porciani}), but it is
fair to say that an accurate estimate of this effect eludes us.  For
low-redshift surveys, for example SuperCOSMOS and the Sloan
spectroscopic galaxy sample, where the median redshift is $z_m \sim
0.1$, the correlation of ellipticities due to intrinsic alignment is
far greater than the expected lensing signal.  As the survey depth
increases, galaxies at fixed angular separation are distributed over
wider ranges of redshifts, leaving only a small proportion of the
galaxy pairs close enough for tidal interactions to correlate the
ellipticities.  This reduction in the total angular intrinsic
alignment signal, combined with an increasing line-of-sight distance
boosting the lensing signal, leaves a smaller intrinsic
alignment contamination for deeper surveys.  The majority of studies
find that for surveys with depths $z_m \sim 1$ the estimated
contamination contributes to up to $\sim 10\%$ of the lensing signal,
although \scite{Jing} has argued that the contamination could be much
higher.  As cosmic shear measurements become more accurate these low
levels of contamination cannot be considered negligible.

If an accurate estimate of the intrinsic ellipticity correlation strength
did exist, then it would be possible to subtract it, leaving
the correlation induced purely by lensing.  In the absence of a good
estimate for the intrinsic  correlation function, it is obvious that
one can improve upon a straightforward ellipticity correlation by
down-weighting galaxy pairs close in redshift and angular sky
separation.  This can be done at the expense of increasing the shot noise
contribution from the distribution of individual galaxy ellipticities.

In this work, we deduce the optimal pair weighting for an arbitrary
survey, given accurate information about the galaxy distances.
We use these results on the Sloan spectroscopic survey design
to show that the optimal weighting scheme can largely remove the
contamination by intrinsic alignments leaving almost pure shot noise. The
shallow depth of this spectroscopic sample still prevents its use for
weak lensing studies as the remaining shot noise exceeds the
expected weak lensing signal.

For multi-colour surveys with photometric redshift information we
propose a semi-optimised procedure of excluding pairs which are likely
to be close in three dimensions.  Comparison with the optimal method
shows this can be very effective.  We apply this technique to the
Sloan photometric survey (SDSS) with $z_m \sim 0.2$, and three deeper
surveys, the Red-Sequence Cluster Survey (RCS) with $z_m \sim 0.56$,
\cite{HYGBHI}, a sample of the COMBO-17 survey with $z_m \sim 0.6$,
\cite{C17} and the Oxford Dartmouth Thirty survey (ODT) with $z_m \sim
1.0$, \cite{Allen}.  We find that even with fairly inaccurate
photometric redshifts it is possible to reduce the contamination from
intrinsic alignments significantly.  For the photometric SDSS, the
improvement is dramatic, enabling it to be used as a weak lensing
survey.  For the deeper surveys we show that using this weighting
scheme, intrinsic alignment contamination can be reduced by several
orders of magnitude.

The effect of the non-uniform weighting on the weak lensing shear
correlation function is calculated.  We find that galaxy pair
weighting slightly reduces the lensing signal, the reduction dependent
on the photometric redshift accuracy and the survey depth.

This paper is organised as follows. In $\S$\ref{sec:theory} we briefly
describe related weak lensing theory. In \S\ref{sec:method} we derive
the optimal pair weighting for a spectroscopic survey.  We discuss the
results of pair weighting for the Sloan spectroscopic sample in
$\S$\ref{sec:res_sloan}, comparing HRH and Jing intrinsic alignment 
models.  In $\S$\ref{sec:photoz} we propose an alternative scheme for
multi-colour surveys with photometric redshift information and present
the results in $\S$\ref{sec:results}.  These results are compared with
the expected weighted lensing signal derived in the Appendix.  In
$\S$\ref{sec:conc} we discuss the results and the implications for the
design of future weak lensing surveys.

\section{Theory}
\label{sec:theory}
In the weak lensing limit the ellipticity of a galaxy, $(e_1,e_2)$, is
an unbiased estimate of the shear $\gamma$, the stretching and
compression of the galaxy image caused by the gravitational lensing of
its light.  Ellipticity is defined by approximating each galaxy as an
ellipse with axial ratio $\beta$,
at position angle $\phi$, measured counterclockwise from the
x-axis.
\be
\left(
\begin{array}{c}
e_1 \\
e_2
\end{array}
\right) =
\frac{\beta^2 - 1}{\beta^2 + 1}
\left(
\begin{array}{c}
\cos 2\phi \\
\sin 2\phi
\end{array}
\right) \ee The complex shear $\gamma$ is related to the average
galaxy ellipticity by $\gamma \simeq \langle e \rangle/2 \equiv
\langle e_1 + i e_2 \rangle / 2$.  A
useful statistic which can be used to constrain the matter power
spectrum is the shear correlation function, $\langle \gamma \gamma^*
\rangle_{\theta}$, for galaxies separated by angle $\theta$.  For
randomly-orientated galaxies in the absence of weak lensing, a
correlation of galaxy ellipticities would yield a zero result at all
angular scales.  The presence of weak lensing introduces
coherent shearing distortions, producing ellipticity or shear
correlations which are related to the nonlinear mass power spectrum
$P_\delta$ via
\begin{equation}
\langle \gamma \gamma^* \rangle_\theta =
\frac{1}{2\pi}\int dk \,k \,P_\kappa(k) \, J_0(k\theta),
\label{eqn:ggt}
\end{equation}
where $J_0$ is the zeroth Bessel function of the first order and
$P_\kappa(k)$ is the convergence power spectrum at wave number $k$,
\begin{equation}
P_\kappa(k) = \frac{9 H_0^4 \Omega_m^2}{4c^4} \int_0^{w_H} dw \,
\frac{g^2(w)}{a^2(w)} \, P_\delta \left( \frac{k}{f_K(w)},w \right),
\end{equation}
$f_K(w)$ is the comoving angular diameter distance out to a comoving
radial geodesic distance $w$, and is defined in the Appendix; $a(w)$
is the dimensionless scale factor, $H_0$ is the Hubble parameter and
$\Omega_m$ the matter density parameter.  The second argument of $P_\delta$
allows for time-evolution of the power spectrum.  $g(w)$ is a
weighting function locating the lensed sources,
\begin{equation}
g(w) = \int_w^{w_H}\, dw'\ \phi(w')
\frac{f_K(w'-w)}{f_K(w')}.
\label{eqn:W}
\end{equation}
$\phi(w(z))$ is the redshift distribution or selection function, and
$w_H$ is the horizon distance, \cite{Bible}.

In addition to shear correlations produced by lensing, several groups
have found evidence for an intrinsic alignment component.  This means
that to some degree, galaxies are not randomly orientated.  In the
absence of intrinsic correlations, the shear correlation function $\langle
\gamma \gamma^* \rangle_\theta$ can be estimated from a catalogue of
galaxy shapes through estimates of the ellipticity correlation
function
\begin{equation}
\widehat{\langle e_a e_b^*\rangle}_\theta = Real\left[\frac{
\sum_{ab} W_{ab}\, {\rm e}_a(\bx) \, {\rm e}_b^*(\bx +
\bm{\theta})} {\sum_{ab} W_{ab}}\right] \label{eqn:ctheta}
\end{equation}
where $e_a$ is the ellipticity of galaxy $a$.
In practice the pair sum is taken over galaxies in a
small range $\Delta\theta$.

Typically the weight assigned to each pair of galaxies, $W_{ab}$,
is the product of the uncertainty in each galaxy shape measurement.
In the next section we derive an additional optimal pair weight which,
in effect, down-weights galaxy pairs close enough to contribute to a
contaminating intrinsic alignment signal, enabling the ellipticity
correlation estimator to be directly related to the lensing shear
correlation function.

\section{Optimal Pair Weighting for Spectroscopic Surveys}
\label{sec:method}

The ellipticity correlation function, for pairs separated by an angle
$\theta$ is given by equation~\ref{eqn:ctheta}.  This
estimate includes an uncertain contribution from intrinsic
alignments, $I_{ab}$.  The estimate
of the lensing shear correlation function $E[\gamma\gamma^*;\theta]$ is
therefore
\be E[\gamma\gamma^*;\theta]  = \frac{1}{4}{\sum_{ab}W_{ab}\left(e_a
e_b^*-I_{ab}\right)  \over \sum_{ab} W_{ab}}
\label{LensingC}
\ee
The amplitude of the intrinsic correlation is controlled by the
comoving distance between the galaxy pair.  It is only significant for
galaxies closer than a few tens of Mpc at most,
in which case the comoving separation is given by
\be
R_{ab}^2 \simeq
(w_a-w_b)^2 + \left[\f\left({w_a+w_b\over 2}\right)
\right]^2 \theta^2.
\ee
In principle, the intrinsic ellipticity correlation depends on the
orientation of the galaxy pair in three dimensions, and also  on the
redshifts of the galaxies. However, studies of the intrinsic effect
average over the orientation, and there seems to  be little evolution
with redshift.  Thus we assume we  have an estimate of an
average correlation function which depends only on the separation $R$
of the two galaxies:
\be
\eta_I(R) \equiv \langle e(\bx) e^*(\bx+{\bf
R}) \rangle_I.
\ee
where the average is over galaxy positions $\bx$, assumed to be known
precisely. In practice, peculiar velocity distortions and
spectroscopic redshift uncertainties lead to errors of order
$\Delta z \sim 0.001$, which we will neglect.  This is justified because, as
we show in section 6, surveys with photometric redshift errors much larger
than this can have intrinsic signals removed virtually as effectively as the
idealised example treated here.
$\eta_I(R)$ can be calculated approximately, from numerical
simulations (HRH, \pcite{CM00,Jing}) or analytic approximation
\cite{CKB01,CNPT01,HZ02,LP01,Porciani}.  The published results do differ
by more than an order of magnitude, and this uncertainty has to be
considered when choosing which $\eta_I(R)$ model to use in the
following pair weighting scheme.
The preferred strategy when applying this technique
is to assume the largest theoretical model for $\eta_I(R)$ to ensure
that all feasible contamination from intrinsic alignments is removed.
This is discussed further in section 6.

To calculate the angular intrinsic contribution to the weak lensing
signal, $\eta_I(R)$ is averaged over galaxy pairs separated by
$\theta$ with a redshift distribution characteristic of the depth of
the survey.  This can be compared to the signal measured from low-redshift
surveys such as the SuperCOSMOS survey \cite{BTHD02}, where
ellipticity correlations are dominated by intrinsic alignments.  These
studies show very rough agreement of the amplitude of the effect, but
the problem is sufficiently difficult that a definitive accurate study
has yet to emerge.  To quantify our uncertainty on the intrinsic
correlations, we assume the intrinsic alignment prediction has a
fractional error $f$ and thus a variance
\begin{equation}
\sigma_{IA}^2 = f^2 \langle I_{ab}\rangle^2
\end{equation}
where $\langle I_{ab}\rangle$ is a weighted average.
and we will conservatively take $f \simeq 1$ in what
follows.  The error in the estimator of the
lensing correlation function, equation~\ref{LensingC},
has two main sources: uncertainty in the
intrinsic correction, and shot noise from the intrinsic  ellipticity
distribution.  If the variance of Real($e_a e_b^*$) is  $\sigma_{\rm pair}^2$,
then the shot noise error is
\begin{equation}
\sigma_{\rm SN}^2 =
{\sum_{ab}W_{ab}^2 \sigma_{\rm pair}^2\over  \left(\sum_{ab}
W_{ab}\right)^2}.
\end{equation}
In the limit of an infinite number of galaxies, $\sigma_{\rm SN}$
tends to zero.

Considering the shot noise error and intrinsic
alignment error only, the total error on the lensing correlation
function is then given by
\be  \sigma_{L}^2 =  {\sum_{ab}W_{ab}^2
\sigma_{\rm pair}^2 + \left(\sum_{ab}W_{ab} f  I_{ab}\right)^2\over
\left(\sum_{ab} W_{ab}\right)^2}
\label{Variance}
\ee
We minimise this variance subject to  $\sum_{ab} W_{ab}=$
constant. The weights for a galaxy pair $p = \{a,b\}$  then satisfy
\be
W_p \sigma_{\rm pair}^2 + f^2 I_p \sum_q W_q I_q  = {\lambda'\over 2}
\ee
where $\lambda'$ is a Lagrange  multiplier.  If we define a matrix
\be
M_{pq} \equiv J_p J_q
\ee
where
\be  J_p \equiv {f I_p\over \sigma_{\rm pair}}
\ee
then $W_p$ satisfies
\be
W_p +M_{pq}W_q = \lambda U_p
\ee
where $\lambda$ is an unimportant constant which we
will set later by  choosing the weights of uncorrelated pairs
($J_p$=0) to be  unity.  $U_p$ is a `unit' vector consisting entirely
of ones.  The solution is
\be
{\bf W} = ({\bf I}+{\bf M})^{-1} {\bf U}.
\label{ww}
\ee
Because of the form of {\bf M}, the inverse can be  computed, by
expanding $({\bf I}+{\bf M})^{-1} = {\bf I}-{\bf  M}+{\bf
M}^2-\ldots$, with elements $[({\bf I}+{\bf  M})^{-1}]_{pq} =
\delta_{pq} + J_p J_q (-1+\Lambda_2-\Lambda_2^2+  \ldots)$, where  \be
\Lambda_2 \equiv \sum_r J_r^2.
\ee
Hence
\be
[({\bf I}+{\bf M})^{-1}]_{pq} = \delta_{pq} - {J_p J_q \over  1+\Lambda_2}.
\label{inv}
\ee
The weight is then obtained from  equation
(\ref{ww}), equivalent to summing (\ref{inv}) over pairs  $q$:
\be
W_p = 1 - {J_p \Lambda_1\over 1+\Lambda_2}
\ee
where  $\Lambda_1 \equiv \sum_r J_r$ and we have chosen $\lambda=1$.
This is the optimal weighting scheme if distances to the galaxies are
accurately known.

\subsection{Reduction in variance}

The variance in the lensing signal is given by equation
(\ref{Variance}), which  may be written
\be  \sigma_L^2 = \sigma_{\rm
pair}^2 {\left[\sum_p W_p^2 + \left(\sum_p W_p J_p\right)^2
\right]\over \left(\sum_p W_p\right)^2}.
\label{Variance2}
\ee
For comparison, with equal weighting of $N$ galaxy pairs,  the variance is
\begin{equation}
\sigma_L^2 =
{\sigma_{\rm pair}^2 \over N^2}\left( N+\Lambda_1^2\right)  \qquad
{\rm [equal\ weighting]}.
\end{equation}
For optimal weighting,
\begin{equation}
\sigma_L^2 = \sigma_{\rm pair}^2{
\left(1+\Lambda_2\right)  \over N
\left(1+\Lambda_2\right)-\Lambda_1^2} \qquad {\rm [optimal]}.
\end{equation}
The ratio of the optimal
variance to the equal weighting variance is shown in Fig. 1  as a
function of $\lambda_1 \equiv \Lambda_1/\sqrt{N}$ and $\Lambda_2$, and
is given by
\begin{equation}
{\rm Ratio} = {(1+\Lambda_2)\over
(1+\lambda_1^2)(1+\Lambda_2  -\lambda_1^2)}.
\label{eqn:ratio}
\end{equation}
Note that
$\lambda_1^2\le \Lambda_2$, which can be shown by considering the
positive quantity
$\sum_p(J_p-  \langle J_p \rangle)^2$.

\begin{figure}
\centerline{\psfig{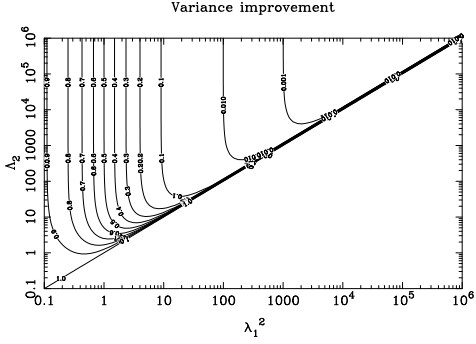}}
\caption{Ratio of optimal variance to the equal weighting variance.\label{Ratio}}
\end{figure}
The sums over galaxy pairs may be estimated by integrals over the mean
number density  of the survey, assuming that we have accurate comoving
galaxy distances which could come from a spectroscopic
redshift survey such as the Sloan spectroscopic galaxy sample.
Redshifts can be converted into distance coordinates $w$ using the
approximate formulae in \scite{Pen99}.

In order to set the weights, we need to compute
\ba
\Lambda_n & = &
\left({f\over \sigma_{\rm pair}}\right)^n \sum_{a,b}\langle I_{ab}
\rangle ^n\nn & \simeq & 2\pi \theta\Delta\theta\Delta\Omega
\left({f\over \sigma_{\rm pair}}\right)^n \int dw_a dw_b \nn & &
\eta_I^n(R_{ab}) \phi_{w}(w_a) \phi_{w}(w_b)
\left[1+\xi(R_{ab})\right]
\ea
where the selection function $\phi_{w}(w)$ is the number of objects at
distance $w$ included in the survey, and $\xi$ is the galaxy two-point
correlation function.  $\phi_w$ is related to the redshift selection
function $\phi_z$ by $\phi_w dw = \phi_z dz $ and we use the redshift
distribution of the form.
\begin{equation}
\phi_z(z) \, \propto \, z^2 \exp \left[
-\left(\frac{z}{z_0}\right)^{1.5} \right]
\end{equation}
where the median redshift of the survey is $z_m \approx 1.4 z_0$.
This is normalised by $\int dz \,\phi_z(z) = n_{\rm str}$, where
$n_{\rm str}$ is the number of galaxies per steradian in the
survey. Note that for $n=0$ we calculate $N$, the expected number of
pairs at separation between $\theta$ and $\theta+\Delta\theta$.  For
$\sigma_{\rm pair}$, we assume that the dispersion in ellipticity is
$\sigma_e \sim 0.2$, \cite{Hudson}, giving $\sigma_{\rm pair} \sim 0.04$.

Typically, for angular scales less than $\theta \simeq 15$ arcminutes,
we find that $\lambda_1^2 \ll \Lambda_2$.  The ratio,
equation~\ref{eqn:ratio}, is then simply $\lambda_1^{-2} $, and the
optimally-weighted error is almost pure shot-noise $\sigma^2_L \approx
\sigma^2_{\rm pair} / N$.  Changing $\Delta\theta$ predominantly
alters $N$ and hence the amplitude of the shot noise component.

Note that the weights are not constrained to be positive and in
extreme cases, if $J_p$ is large, can be negative.  The
reason for this is that the contribution to the error from intrinsic
alignments is assumed to be proportional to the intrinsic alignment
signal itself.  If this is the dominant error term, it is possible
that one can have a net gain by reducing this term through giving
some pairs negative weight.   This behaviour is, however, rare.

\section{Results for the Sloan Spectroscopic Sample: Comparison of HRH
and Jing intrinsic alignment models}
\label{sec:res_sloan}
\begin{figure}
\centerline{\psfig{file=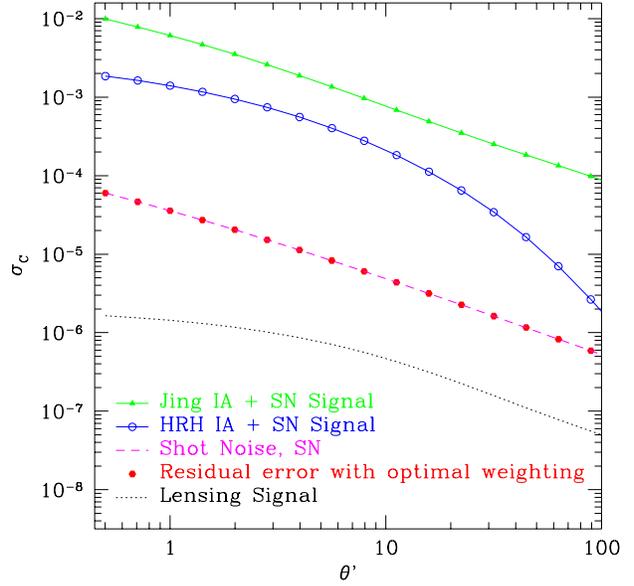,width=8.5cm,angle=0,clip=}}
\caption{Reduction in the error from intrinsic alignments and
shot noise for the Sloan spectroscopic sample.  The upper curves
show the error in the shear correlation function for two
intrinsic alignment models, assuming all galaxy pairs are
weighted equally. Note that these curves essentially combine a
statistical error 
(shot noise) with a systematic contamination (intrinsic alignments).
Triangles show the Jing model and circles HRH's
spiral model. The error with optimal weighting is shown as filled
circles, and has reduced both intrinsic alignment model signals
close to shot noise (shown dashed). Note that the increase
in shot noise is negligible due to the small proportion of downweighted
pairs in the optimal weighting scheme. The expected weak
gravitational lensing correlation function for a $\Lambda$CDM
model, (dotted), is still dominated by shot noise for this
shallow $z_m = 0.1$ sample.  } \label{pic:SDSSsp}
\end{figure}

Covering $\pi$ steradian the Sloan Digital Sky Survey, SDSS, is a
wide, shallow survey.  The spectroscopic galaxy sample has a median
redshift of 0.1 with redshift measurements known to an accuracy of
$\Delta z = 0.0001$ \cite{SEDR}.  Figure~\ref{pic:SDSSsp} shows the
reduced intrinsic alignment error calculated using the optimal
weighting scheme, compared to the error introduced to the lensing
signal by an unweighted HRH intrinsic alignment signal and an
unweighted Jing intrinsic alignment signal at a redshift of 0.1.  The
minimised signal for both intrinsic alignment models is close to pure
shot noise with residuals of less than 1\%.

The published differences between the HRH and Jing models come
from from the higher resolution N-body simulation used by Jing,
and differences in the choice of $\xi(R)$.  Jing finds that the
spatial ellipticity correlation of the halos, $\eta_I(R)$,
derived for the same minimum dark matter halo mass, is higher by a
factor of 2, compared to the elliptical HRH model.  This model
assumes that the ellipticity of the galaxy is the same as that of
the parent halo.  It is sensitive to the accuracy of the halo
shape measurement and is therefore better determined in the higher
resolution simulations.  A second difference, in the angular
intrinsic ellipticity correlation, arises since Jing uses a
larger two-point correlation function, measured from the
simulation itself, whereas HRH assumed $\xi(R) = (R\, / \,5h^{-1}
{\rm Mpc})^{-1.8}$.  HRH also used a model for spiral galaxies,
where the galaxy is modelled as a thin disk, perpendicular to the
angular momentum vector of the parent halo. The spiral model has
significantly lower correlations at spatial separations $R_{ab}
\gs 5 h^{-1} {\rm Mpc}$.

Except where otherwise stated, we use the HRH spiral model throughout
this paper. $\eta_I(R)$, summed over $e_1$ and $e_2$, is fitted by an
exponential: $\eta_I(R) = 0.012 \exp(-R/[1.5 h^{-1} Mpc])$.  We should
point out that fitting an exponential model could underestimate $\eta_I(R)$
on large scales $R>10 {\rm h}^{-1}{\rm Mpc}$, where the signal-to-noise
is low.  This would then lead to an underestimation of the intrinsic
signal on angular scales $\theta \ge 10'$.  If this were the case, however,
the results would be inconsistent with the lack of detection of a
B mode signal on these scales found by \scite{HYG02}, 
discussed further in section 6. Figure~\ref{pic:SDSSsp} shows the
contamination expected from this exponential HRH model and the
larger signal found by \scite{Jing}, assuming the same minimum halo
mass as HRH ($6.9 \times 10^{11} {\rm M}_\odot$).  We have, however,
used the power-law two-point correlation function to convert
the Jing $\eta_I(R)$ fitting formula into an angular signal.
Increasing the minimum halo mass increases the unweighted Jing signal,
but does not greatly affect the minimised signal leaving residuals at
less than 3\% of the shot noise.

As the Sloan spectroscopic sample is so shallow, the expected lensing
signal is negligible preventing this spectroscopic sample from being
used as a weak lensing data set.  It is however an interesting test of
the method showing that the optimal weighting scheme will almost
completely remove both the HRH and the stronger Jing model for
intrinsic alignments.  We will therefore proceed with the
HRH spiral model for intrinsic alignment, as the method has been shown
to remove successfully any amplitude of intrinsic alignment provided the
signal is significant only for galaxy pairs closer than a few tens of
Mpc.

\section{Using Photometric Redshifts}
\label{sec:photoz}

The method detailed in section~\ref{sec:method} is only feasible for
spectroscopic surveys with essentially known galaxy distances.  
More typically we
have good estimates of galaxy distances from photometric redshifts
with associated errors.  These errors are much larger than the scale
over which the intrinsic alignments are correlated.  We must therefore
define a new weighting scheme for galaxy pairs dependent on their
estimated redshifts alone.  Those weights are then transformed into an
average weight assigned to a pair at true redshift separation to
calculate the residual angular intrinsic contribution.

\begin{table}
\begin{center}
\begin{tabular}{c|c|c|c|c|c}
Survey & size str & $z_m$ & $\Delta_z$ & \\ \hline  Sloan
spectroscopic sample
& $\pi$ & 0.1 & 0.0001 \\  SDSS photometric & $\pi$ & 0.2 & 0.025 \\
COMBO-17 & 0.0005 & 0.6 & 0.03 \\  RCS BVRz' & 0.015 & 0.56 & 0.3 \\  ODT
UBVRIz' & 0.005 & 1.0 & 0.2 \\
\end{tabular}
\end{center}
\caption{ Survey parameters for SDSS, COMBO-17, RCS and ODT}
\label{table}
\end{table}

Using the minimum lensing error derived in
section~\ref{sec:method} as a benchmark we propose a simpler
approach to derive the weightings for galaxies at estimated
distances.  Applying the accuracy of the Sloan spectroscopic
sample, $\Delta_z = 0.0001$, to this alternative method we come
very close to reproducing the previously derived minimum.  This
shows that our alternative method, in the limit of accurate
photometric redshift information, is close to the optimal
result.  As the errors in the distance measurements increase, our
ability to down-weight close galaxy pairs correctly decreases, and
the residual intrinsic contamination increases.

\begin{figure}
\centerline{\psfig{file=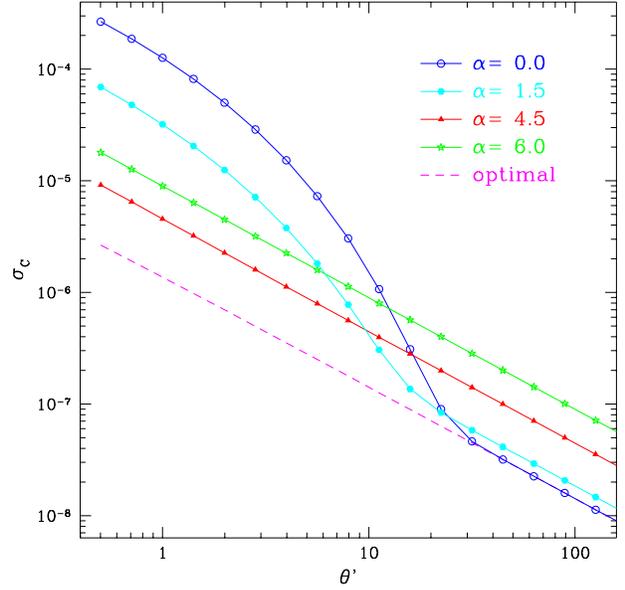,width=8.5cm,angle=0,clip=}}
\caption{Residual error expected after removing pairs within $\alpha \Delta_{z}$
of each other for the ODT survey.  $\Delta_{z}$ is the root mean
square photometric redshift error.}
\label{pic:alpha}
\end{figure}

For a pair of galaxies with estimated redshift $\hat z_a$ and $\hat
z_b$ we assign a zero weight if $ | \hat z_a - \hat z_b| < \alpha
\Delta_z $ and a weight of one otherwise.  We choose $\alpha$ to
minimise the total error on the shear correlation function.  The
optimum value depends on angular separation, and survey
details. Clearly, this is not the most general weighting scheme, but
we will see later that, with good photometric redshifts, this simple
procedure does almost as well as the theoretical optimum where all
galaxy distances are known.  Assuming that the estimated redshifts are
distributed normally about the true redshifts with Gaussian widths
given by $\Delta_z$, we find that the average weight for a galaxy pair
at true redshifts $z_a$ and $z_b$ is given by
\be
\langle
W_{ab}\rangle = 1 - \frac{1}{2 \sqrt{\pi} } \int dx \, {\rm e}^{-x^2}
\left[ {\rm erf}(y+x) - {\rm erf} (v + x) \right] \ee where \be x
\equiv \frac{\hat z_a - z_a}{ \sqrt{2} \, \Delta_z} \,\,\,\, y \equiv
\frac{z_a - z_b + \alpha\Delta_z }{\sqrt{2} \, \Delta_z} \,\,\,\, v
\equiv \frac{z_a - z_b - \alpha\Delta_z }{\sqrt{2} \, \Delta_z}
\ee
The weighted intrinsic component can then be evaluated using the
integral version of equation~\ref{Variance}, setting the value of
$\alpha$ by minimising the total error contribution from the intrinsic
alignments, $\sigma_{IA}$ and the shot noise, $\sigma_{SN}$.

\be
\sigma_{IA} = \frac{ f \int dz_a dz_b   \phi_{z}(z_a)
\phi_{z}(z_b)  \left[1+\xi(R_{ab})\right] \langle W_{ab}\rangle
\eta_I(R_{ab})}{ \int dz_a dz_b   \phi_{z}(z_a) \phi_{z}(z_b)
\left[1+\xi(R_{ab})\right] \langle W_{ab}\rangle}
\ee

\begin{equation}
\sigma_{SN} = \sigma_{\rm pair} \left( \int dz_a dz_b
\phi_{z}(z_a) \phi_{z}(z_b) \left[1+\xi(R_{ab})\right] \langle
W_{ab}\rangle \right)^{-1}
\end{equation}
As $\alpha$ increases the probability of removing all close galaxy
pairs increases and the intrinsic contribution decreases to zero.
This means that the total galaxy count decreases and the shot noise
increases, hence there is an optimum value of $\alpha$.  As $\theta$
increases, shot noise begins to dominate the intrinsic alignment
error, and a progressively lower value of $\alpha$ is favoured.  This
is shown in figure~\ref{pic:alpha} where the reduced intrinsic
contamination is plotted for varying values of $\alpha$ for a low
accuracy photometric redshift survey, for example the ODT.

The optimum values of $\alpha$ are dependent on survey size, depth,
photometric redshift accuracy and the choice of $\eta_I(R)$.

\section{Application of semi-optimised
weighting scheme to multi-colour survey designs}
\label{sec:results}

We consider four multi-colour surveys with varying depth, angular size
and photometric redshift accuracy: Sloan photometric survey, SDSS,
COMBO-17, the RCS survey and the ODT survey.  Table~\ref{table}
details the relevant survey parameters.  The uncertainty in the
redshift depends mainly on the number of colours known for each
object.  Photometric redshift errors for SDSS assumes a
neural-network-based photometric redshift estimator
\cite{Nets1,Nets2}.  For the RCS and ODT we assume the use of hyper-z
\cite{hyperz}.  The full COMBO-17 survey has imaging to $R\sim 25.5$,
but colour information in 17 bands (producing very accurate
photometric redshift estimates) down to a limiting magnitude of $R
\sim 24$.  It is this magnitude-limited sample of the full COMBO-17
weak lensing survey which we will consider.

\begin{figure}
\centerline{\psfig{file=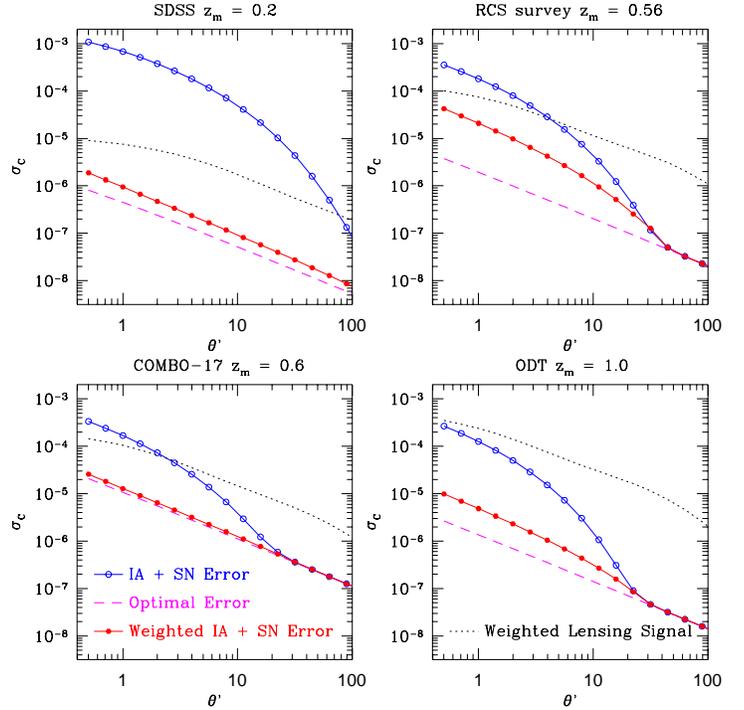,width=10.0cm,angle=0,clip=}}
\caption{Reduction in the error from intrinsic alignments and
shot noise for photometric SDSS, RCS, COMBO-17 and ODT.  The
semi-optimal weighting, (filled circles), has reduced the
unweighted HRH intrinsic alignment error (circles) to well below
the expected amplitude of the weighted weak lensing shear signal,
(dotted). The effect of semi-optimal weighting can be compared to
the optimally weighted error, (dashed), attainable with
spectroscopic redshifts.   } \label{pic:all}
\end{figure}

\begin{figure}
\centerline{\psfig{file=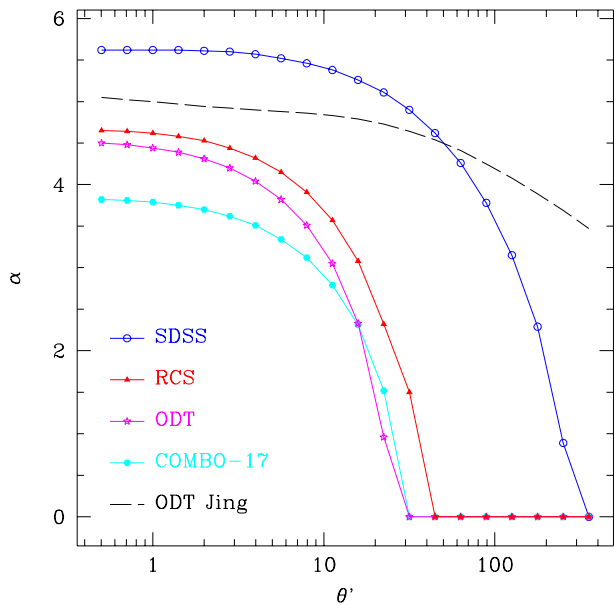,width=8.5cm,angle=0,clip=}}
\caption{Optimum values of $\alpha$ for the four different surveys,
assuming the HRH spiral model for intrinsic alignments.
The dashed line shows the optimal values of $\alpha$ for the ODT
survey, assuming the higher intrinsic signal found by Jing (2002).
At the expense of marginally-higher residual shot noise, this
conservative pruning of the pairs virtually guarantees almost complete
removal of the intrinsic alignment contamination.}
\label{pic:best}
\end{figure}

Figure~\ref{pic:all} shows the correlation signal we expect to find
from HRH-derived intrinsic alignments and the best reduction that can
be achieved with this method, 
with photometric redshift information at the accuracy
appointed for each survey.  These have been calculated using the
values of $\alpha(\theta)$ shown in figure~\ref{pic:best}.  The dashed
line in Figure~\ref{pic:all} shows the optimal improvement in the
error obtainable if accurate distances were known, using the method
detailed in section~\ref{sec:method}.  This acts as a useful benchmark
to see how close the semi-optimal method for multi-colour surveys gets
to the ideal.  The intrinsic alignment signals can be compared to the
expected weighted lensing signal, equation~\ref{eqn:wls}, for each
survey.  This is derived from a nonlinear CDM mass power spectrum with
$\Omega_{m} = 0.3$, $\Omega_{\Lambda} = 0.7$, $\Gamma = 0.21 $,
$\sigma_8 = 0.9$, calculated using the `halo model' fitting formula
\cite{Rob}.  For the pair weighting proposed, this signal is slightly
lower than the unweighted lensing signal (equation~\ref{eqn:ggt}),
(see Appendix).

The true amplitude of the intrinsic alignment signal is rather
uncertain; the correlation signal from intrinsic alignments could move
up, as argued by \scite{Jing}, or down, as argued, for example, by
\scite{CNPT01}, and by up to an order of magnitude.
Figure~\ref{pic:best} also shows the optimum $\alpha(\theta)$ if the
intrinsic alignment signal is as high as found by \scite{Jing}, for
the ODT survey.  Due to the presence of intrinsic correlations at
larger galaxy pair spatial separations, it is necessary to remove more
pairs at larger angular scales.  At the expense of a higher residual
shot noise, this more-or-less
guarantees removal of any plausible intrinsic alignment signal,
leaving errors at less than 3\% of the weighted ODT lensing
signal.  This is probably the preferred strategy until the intrinsic
signal is better determined, but it should be noted that an intrinsic
alignment signal of the amplitude found by \scite{Jing} is
incompatible with current weak lensing detections, \cite{vWMT}.

Using an HRH model for intrinsic alignments we find that the SDSS is
dominated by intrinsic alignments and that the weak lensing signal
from the two middle-depth surveys, RCS and COMBO-17, will be strongly
contaminated at angular scales less than 10 arcminutes.  Interestingly
\scite{HYG02} show an E/B mode decomposition for the RCS survey which
shows a significant B mode at angular scales less than 10 arcminutes.
The presence of a B mode can be due, on small angular scales,
to source clustering, \cite{SchvWM02}, or it can imply systematics
within the data and/or the presence of intrinsic galaxy alignments
\cite{CNPT02}.

The application of the weighting scheme produces encouraging results,
the most startling of which comes from the SDSS.  Due to the wide sky
coverage and accurate photometric redshift information it is possible
to reduce the intrinsic alignment signal from well above, to well
below the expected lensing signal. This enables, in principle, the
extraction of a lensing signal at angular scales, $\theta > 1'$, with
errors less than $10\%$.  We see similar reductions for the other
three surveys, with the success of the reduction dependent on the
accuracy of the redshift estimates.

\section{Conclusions and implications for future weak lensing surveys}
\label{sec:conc}
In this paper we have shown how distance information can be used to
reduce the contamination of the weak gravitational lensing shear
signal by the intrinsic alignments of galaxies.  Our principal finding
is that the level of intrinsic ellipticity correlation in the shear
correlation function can be reduced by up to several orders of
magnitude, by down-weighting appropriately the contribution from pairs
of galaxies which are likely to be close in three dimensions.

For spectroscopic surveys we have derived an optimal galaxy pair
weighting which reduces the contamination to a negligible component,
leaving almost pure shot noise.   Application of the technique to the
relatively shallow Sloan spectroscopic galaxy sample reduces the error
by up to two orders of magnitude, but still leaves the lensing signal
undetectable, dominated now by shot noise.

For multi-colour surveys with photometric redshift estimates  we have
proposed a partially-optimised method, which removes pairs with close
photometric redshifts from the computation of the shear correlation
function. We find that with accurate photometric redshifts this simple
method is almost as effective as the fully optimised method.  We show
that, even with relatively crude photometric redshift estimates, the
contamination by intrinsic alignments can be significantly reduced.

Similar conclusions were
independently drawn by \scite{KingSch02} who simultaneously proposed a
weighting scheme based on photometric redshifts.
Our technique however requires the assumed knowledge of a model for the
intrinsic alignments.  Provided one is conservative, assuming the
largest feasible model, this weighting scheme will reduce the
true intrinsic alignment contamination to lensing correlation signals
to a negligible level.

We have applied the method to four multi-colour survey designs.  The
shallow photometric SDSS survey with $z_m \sim 0.2$,  shows the most
dramatic of improvements.  The intrinsic alignment signal is expected
to dominate completely the weak lensing shear signal from a survey of
this depth, but with judicious removal of pairs within $\sim 0.14$ in
estimated redshift, the error from shot noise and intrinsic alignments
is reduced to only $\sim 10\%$ of the expected shear signal.  This
opens up the possibility of using the SDSS for future cosmic shear
studies. With deeper surveys, such as the RCS, COMBO-17 and the ODT
survey, intrinsic alignments in the weak lensing shear signal are
significantly reduced. With the highly accurate photometric redshifts
of COMBO-17, the reduction is close to optimal, reducing the largely
systematic $\sim 220\%$ error at small angular scales to a much
smaller and largely statistical error of $\sim 17\%$ of the expected
lensing shear correlation. This limiting error is predominantly 
shot noise which will decrease as the survey size grows. The ODT and
RCS have less accurate distance estimates but even with current
photometric redshift estimates, which undoubtedly will improve with
time, the reduction is quite significant.  For the wide RCS, the 130\%
contamination at angular scales $\theta \sim 2'$, is reduced to an error
$\sim 20\%$. For deep multi-colour surveys such as the ODT, the intrinsic
alignment signal is almost completely removed, leaving noise at a
level of $\leq 2$\% of the lensing signal.  

The aim for weak lensing surveys to date has been to go as deep and as
wide as possible.  With an increase in depth comes an increase in the
expected lensing signal and for purely geometrical reasons we also see
a decrease in intrinsic alignment correlations.  To reduce shot noise,
the error introduced by the intrinsic ellipticity distribution of
galaxies, surveys can go wide and/or deep.  This increases the number
of lensed sources, in the limit of an infinite number of galaxies the
shot noise goes to zero.  One additional source of error so far
unmentioned is cosmic variance which, for small area surveys, can
dominate on scales larger than a few arcminutes, \cite{SchvWKM02}.
This can be reduced by sampling different areas of the sky and
combining to produce a wide field survey.
Note that downweighting close galaxy pairs affects this random
error little, whilst effectively removing the systematic errors from
intrinsic contamination.

As the size of future lensing surveys
increase, control of systematic errors becomes increasingly important.
Intrinsic alignment contamination is a potentially large source of
error, even if it is at a level much lower than assumed here.
This paper has shown that with the addition of accurate photometric
redshift information, the presence of intrinsic alignments can be
effectively removed.  \scite{HYGBHI} have shown that it is possible to
extract a low weak lensing signal from a survey with a median redshift
of 0.56 and argue that using this shallower survey enables more
accurate star-galaxy separation and provides a fairly well determined
redshift distribution for the sources, another uncertainty in weak
lensing measurements to date.  The addition of depth information to
weak lensing studies also opens new possibilities for the application of
lensing tomography \cite{Hu02,Hu99}, and reconstruction of
the full 3D cosmological mass distribution \cite{AndyT,HuKeeton}, but
here we have concentrated only on its benefits in reducing intrinsic
alignment contamination.  
In view of the dramatic improvements it offers lensing
signal detection we therefore propose that for future weak lensing surveys,
emphasis should be placed on acquiring multi-colour data
in a wide area.  In the case of restricted telescope time,
moderate depth surveys, $z_m \sim 0.6$, may then become a more attractive
alternative to ultra-deep observations, $z_m > 1.0$, that can often be
limited by seeing.

\section{Acknowledgements}

We are very grateful to Rob Smith for providing us with his
nonlinear power spectrum fitting formula and code.  We thank Yipeng
Jing, Michael Brown and Emily MacDonald for useful discussions and the referee
for helpful comments.

\bibliographystyle{mnras}
\bibliography{ceh}

\onecolumn
\appendix
\section{Theoretical lensing signal for weighted ellipticities}
\label{app}

The effective convergence is (\scite{Bible}, equation 6.18)
\be
\k(\th) = A \int_0^\infty dw g(w)\f(w) {\delta[\f(w)\th,w]\over a(w)}
\ee
where $A=3H_0^2 \Omega_m/(2c^2)$, $g(w)$ is the weighting function
given in equation~\ref{eqn:W}, and $\f(w)$
is the geometry-dependent comoving angular diameter distance,
dependent on the curvature $K$,
\be
f_K(w) =  \left\{
  \begin{array}{ll}
    K^{-1/2}\sin(K^{1/2}w) & (K>0) \\
    w & (K=0) \\
    (-K)^{-1/2}\sinh[(-K)^{1/2}w] & (K<0) \\
  \end{array}\right.\;.
\ee
Note that $w$, the comoving radial geodesic distance, plays two
roles, both as a third spatial coordinate, and as a time evolution
label.

The effective 2D convergence is an integral of the 3D
convergence, and it is this which we will weight, since each
galaxy gives an estimate of the shear, or convergence, in the
weak lensing limit. \be \k(\th) = \int_0^\infty dw \,\phi(w)
\kappa_{\rm eff}(\th,w). \ee We will weight 3D convergence
estimates with a weight function $H(w,w',|\th|)$, to give the
weighted correlation function
\begin{eqnarray}
\langle \k(\th) \k^*(\th')\rangle &=& {1\over Q}\int_0^\infty dw
\int_0^\infty dw'\, H(w,w',|\th-\th'|) \langle \kappa_{\rm
eff}(\th,w)\kappa^*_{\rm eff}(\th',w')\rangle \phi(w)\phi(w')\nn
&=& B \int_0^\infty dw \int_0^\infty dw' \,\phi(w)\phi(w')
H(w,w',|\th-\th'|) \int_0^w dw_1 {\f(w_1)\f(w-w_1)\over
\f(w)a(w_1)}\nn & & \int_0^{w'} dw_2 {\f(w_2)\f(w'-w_2)\over
\f(w')a(w_2)} \langle
\delta(\f(w_1)\th,w_1)\delta^*(\f(w_2)\th',w_2) \rangle
\end{eqnarray}
where $Q\equiv \int_0^\infty\int_0^\infty dwdw'\,H(w,w',\theta)
\phi(w)\phi(w')$ and $B\equiv A^2/Q$. We now assume that the
scale over which the correlations of density are non-zero is
small, in the sense that there is negligible evolution of the
density field over the light-crossing time of the correlation
scale. Writing
\begin{equation}
\delta({\bx}) = \int {d^3\bk\over (2\pi)^3} \delta_\bk(w)
\exp(-i\bk\cdot\bx)
\end{equation}
and using homogeneity of $\delta$ to define the power spectrum
of the density field $P$,
\begin{equation}
\langle \delta_\bk(w) \delta^*_{\bk'}(w') \rangle = (2\pi)^3
P(k,w) \delta^D(\bk - \bk') \ee where $\delta^D$ is a Dirac delta
function.  Hence
\begin{eqnarray}
\langle \k(\th) \k^*(\th')\rangle &=& B \int_0^\infty dw
\int_0^\infty dw' \,\phi(w)\phi(w') H(w,w',|\th-\th'|) \int_0^w
dw_1 \int_0^{w'} dw_2 F(w,w_1)F(w',w_2)\nn & & \int {d^3\bk\over
(2\pi)^3} \exp[-i\bk_\perp\cdot
(\f(w_1)\th-\f(w_2)\th')]\exp[ik_z(w_1-w_2)] P(k,w_1)
\end{eqnarray}
where
$\bk=(\bk_\perp,k_z)$ and \be F(w,w') \equiv {\f(w')\f(w-w')\over
\f(w)a(w')}
\end{equation}
Now we make the standard approximation that the correlation function of
$\delta$ is non-zero only if $w_1$ and $w_2$ are almost equal.  Specifically,
we assume that $F$ and $\f$ don't vary much over this scale.  Thus we set
$\f(w_1)=\f(w_2)$ in the first exponential, and also approximate
$F(w',w_2)\simeq F(w',w_1)$.  The $w_2$ integral then simplifies, since it is
approximately
\begin{equation}
\int_0^\infty dw_2 \exp(-ik_z w_2) = 2\pi \delta^D(k_z)
\end{equation}
and we are left with
\begin{eqnarray}
\langle \k(\th) \k^*(\th')\rangle &=& B \int_0^\infty dw
\int_0^\infty dw' \,\phi(w)\phi(w') H(w,w',|\th-\th'|) \int_0^w
dw_1 F(w,w_1)F(w',w_1)\nn & & \int {d^3\bk_\perp\over (2\pi)^2}
\exp[-i\bk_\perp\cdot \f(w_1)(\th-\th'))]\exp[ik_z w_1]
P(|\bk_\perp|,w_1).
\end{eqnarray}
An angle integration of $P$ gives
\begin{equation}
\int {d^3\bk_\perp\over (2\pi)^2} \exp[-i\bk_\perp\cdot
\f(w_1)(\th-\th'))] P(|\bk_\perp|) = \int_0^\infty {k dk\over
2\pi} P(k,w_1) J_0[k \f(w_1)|\th-\th'|]
\end{equation}
where $J_0$ is the
zeroth Bessel function of the first order. Hence we can write the
correlation function as
\begin{eqnarray}
\langle \k(\th) \k^*(\th')\rangle &=& B \int_0^\infty dw
\int_0^\infty dw' \,\phi(w)\phi(w') H(w,w',|\th-\th'|) \int_0^w
dw_1 F(w,w_1)F(w',w_1)\nn & & \int_0^\infty {k dk\over 2\pi}
P(k,w_1) J_0[k \f(w_1)|\th-\th'|].
\end{eqnarray}
Writing this in a form as close to equation~\ref{eqn:ggt} as
possible, by reversing the order of integration, and noting that
in the weak lensing limit, shear and convergence have the same
statistical properties; $\langle \gamma(\th)
\gamma^*(\th')\rangle = \langle \k(\th) \k^*(\th')\rangle$, we
find:
\begin{equation}
 \langle \gamma(\th) \gamma^*(\th')\rangle =
B \int_0^\infty dw_1 X(w_1,|\th-\th'|)\int_0^\infty {k dk\over
2\pi} P(k,w_1) J_0[k \f(w_1)|\th-\th'|]. \label{eqn:wls}
\end{equation}
where
\begin{equation}
X(w_1,\theta) \equiv \int_{w_1}^\infty dw \int_{w_1}^\infty dw'
\phi(w) \phi(w') F(w,w_1) F(w',w_1) H(w,w',\theta).
\end{equation}
In the equal-weighted case, $H=1$, and $X(w,\phi)$ simplifies to
the product of two equal integrals, each independent of $\phi$
and equal to $g(w)\f(w)/a(w)$.

\begin{figure}
\centerline{\psfig{file=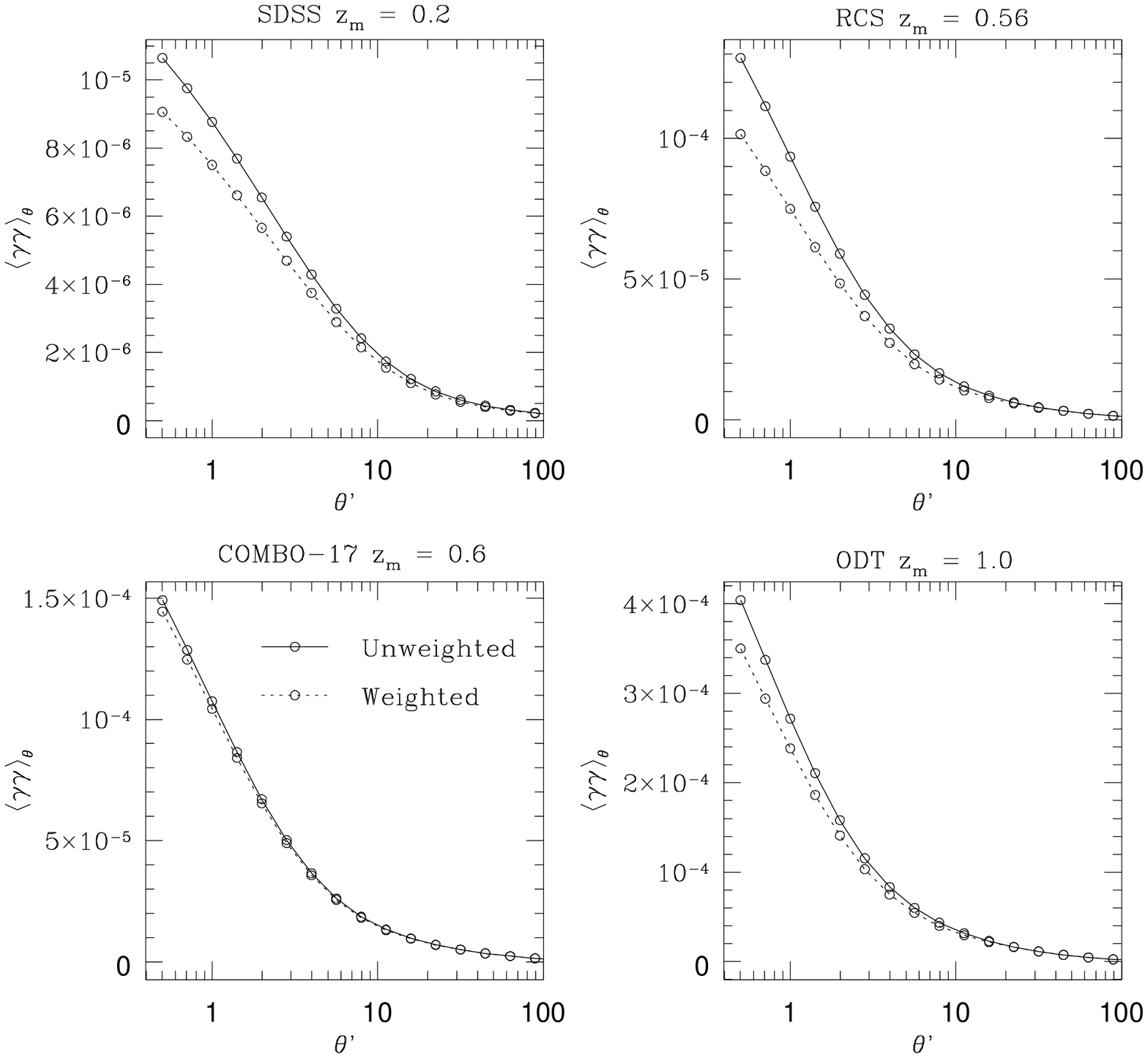,width=10.5cm,angle=0,clip=}}
\caption{Expected weak lensing signal for a $\Lambda CDM$ model with
$\sigma_8=0.9$, for the surveys discussed in the main text.
The solid line shows equal weighting of galaxies; the
dotted line shows the effect of employing the proposed multi-colour
survey weighting scheme.}
\label{weightedsignal}
\end{figure}

The effect of the weighting is shown in Fig.~\ref{weightedsignal}.
For surveys such as COMBO-17 with accurate photometric redshifts, the
lensing signal changes by up to $\sim 3\%$.  For the Sloan photometric
survey, the effect is $\sim 10\%$; the greatest effect is for RCS,
which has the least accurate photometric redshifts.

\end{document}